# Progress on the Electromagnetic Calorimeter Trigger Simulation at the Belle II Experiment

I.S. Lee, S.H. Kim, C.H. Kim, H.E Cho, Y.J. Kim, J.K. Ahn, E.J. Jang, S.K. Choi, Y. Unno and B.G. Cheon

*Abstract*—The Belle II experiment at KEK in Japan has started real data taking from April 2018 to probe a New Physics beyond the Standard Model by measuring CP violation precisely and rare weak decays of heavy quark and lepton. The experiment is performed at the high luminosity SuperKEKB $e^+e^-$ collider with $80 \times 10^{34} cm^{-2}s^{-1}$ as an ultimate instantaneous luminosity. In order to develop and test an appropriate trigger algorithm under much higher luminosity and beam background environment than previous KEKB collider, a detail simulation study of the Belle II calorimeter trigger system is very crucial to operate Belle II Trigger and DAQ system in stable. We report preliminary results on various trigger logics and their efficiencies using physics and beam background Monte Carlo events with a Belle II Geant4-based analysis framework called Basf2.

## I. INTRODUCTION

The Belle II experiment has begun data-taking to probe New Physics beyond the Standard Model by measuring precisely CP violation phenomena and rare decays of beauty, charm quark and tau lepton. In Phase II operation started from April 2018 the SuperKEKB instantaneous luminosity does already reach to the order of $10^{33}$ cm$^{-2}$s$^{-1}$ and we expect to reach to the level of $10^{34}$ cm$^{-2}$s$^{-1}$ in early operation stage of Phase III in 2019. Because the beam background level is 50 ~ 100 times higher than the case of Belle, the robust and flexible trigger system is indispensable to operate the Belle II detector smoothly. For Belle II Data Acquisition (DAQ) suitable to raw data readout we should satisfy several trigger requirements that are high trigger efficiency from $\Upsilon(4S) \to B\bar{B}$ as 100%, the maximum trigger rate of 30 kHz, 5 μs trigger latency, event timing resolution of less than 10 ns accuracy and the minimum event-by-event separation of 200ns, while the trigger efficiencies of low multiplicity physics events sensitive to Dark photon search should be kept as high as possible.

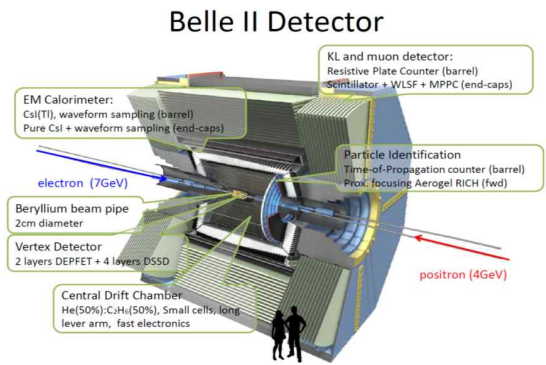

Fig. 1. Belle II detector composed of five kinds of sub-systems.

To meet these requirements, we plan to use the Belle triggering scheme [1] with new technologies. The Belle II trigger system consists of sub-trigger systems and a global decision logic system. A sub-trigger system summarizes trigger information on its subsystem, and sends it to the global decision logic that does combine sub-triggers and generate a final trigger when their trigger conditions are satisfied. In Belle II, we upgraded all system components and connections with new technologies. Each component has a Field Programmable Gate Array (FPGA) so that the trigger logic is flexibly configurable. All trigger data flow along high speed serial links, not parallel (ribbon) cables.

## II. LEVEL 1 TRIGGER SYSTEM

Belle II level 1 hardware trigger system consist of four sub-trigger systems and one global trigger system. All system components were upgraded to handle 60 times larger trigger rate than Belle.

Central Drift Chamber (CDC) trigger system provides momentum, position, charge, and number of tracks information of charged tracks. Electromagnetic Calorimeter (ECL) trigger generates neutral trigger information such as total energy,

This research was supported by Basic Science Research program through the National Research Foundation of Korea (NRF-2016K1A3A7A09005604, 2018R1A2B3003643) and by the research fund of Hanyang University (HY-2009).

I. S. Lee (primary author) and B. G. Cheon (corresponding author) are with the Department of Physics and Research institute for natural sciences, Hanyang University, Seoul, 04763 Korea (e-mail : insoo.lee@belle2.org and: bgcheon@hanyang.ac.kr).

S. H. Kim, C.H. Kim, H.E Cho and Y. Unno are with the Department of Physics, Hanyang University, Seoul, 04763 Korea.

Y. J. Kim and J.K. Ahn are with the Department of Physics, Korea University, Seoul, 02841 Korea.

E.J. Jang and S.K. Choi are with the Department of Physics, Gyeongsang National University, Jinju, 52828 Korea.

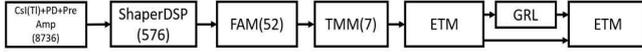

Fig. 2. Software and hardware configuration of the Belle II ECL calorimeter trigger system.

cluster counting, and Bhabha tagging information of electromagnetic particle. In case of Bhabha and γγ physics processes a pre-scale by factor 1/100 is applied. Barrel Particle Identification detector (TOP) trigger delivers precise timing and hit topology information. $K_L^0$ and μ detector (KLM) trigger gives $K_L^0$ and μ particle hit information. Global trigger system deals with CDC-ECL trigger matching and many combined triggers to generate the final trigger bit delivered to Belle II Data Acquisition system (DAQ) [1].

### III. ECL TRIGGER SYSTEM

The basic framework and idea of Belle II ECL trigger system are same as the case of the Belle [1] [2]. The schematic overview of ECL trigger system is shown in Fig. 2. In order to handle higher trigger rate due to high luminosity and beam background level, we have adopted a new trigger scheme that makes the trigger performance more flexible using a readout electronics architecture with Flash Analog-to-Digital Converter (FADC), Field Programmable Gate Array (FPGA) components, and high-speed serial data transfer at 127 Mbps link speed.

Shaper Digital Signal Processing (SDSP) board receives an analog signal through photo-diode (PD) and pre-amplifier attached to the CsI(Tl) crystal counter. The SDSP generates an analog signal with 0.2μs shaping time for trigger purpose. The 16 shaping signals from neighboring 4 × 4, called a Trigger Cell (TC), is combined in SDSP. The TC is the basic unit of the ECL trigger system and a total of 576 TCs are used. The figure 3 shows this TC map. All TC analog signals are sent to FADC Analysis Module (FAM). The FAM does digitize the analog TC signal with FADC and measure each TC energy and timing using minimum $\chi^2$ fit method. Then TC information is sent to Trigger Merger Module (TMM) in which merge the coming TC data and send it to ECL Trigger Master (ETM). The ETM analyze all TC data and generate Bhabha trigger, physics trigger, and event timing. Bhabha trigger is important to luminosity measurement but should be pre-scaled because it is most dominant production process in $e^+e^-$ beam collision. In phase II operation, we use Belle type 2-D Bhabha logic. Using back-to-back topology of Bhabha event, 14 types of θ combinations and their thresholds are decided from simulation study. The θ combination types of 2-D Bhabha logic are summarized in Table I. Main physics triggers are total energy and cluster counting triggers. Total energy should be greater than 1 GeV with a Bhabha veto, and the number of isolated clusters (ICN) should be greater than 3 [3].

TABLE I
14 TYPES OF 2-D BHABHA TAGGING LOGIC

| # | Combination (θ id) | | Energy Threshold (GeV) | |
|---|---|---|---|---|
| 1 | 1 + 2 + 3 | 16 + 17 | 4.0 | 2.5 |
| 2 | 3 | 15 | 4.0 | 2.5 |
| 3 | 4 | 14 + 15 | 4.0 | 2.5 |
| 4 | 5 | 14 + 15 | 4.0 | 2.5 |
| 5 | 4+5 | 14 | 4.0 | 2.5 |
| 6 | 5 | 13 +14 | 4.0 | 2.5 |
| 7 | 5 | 12 +13 | 4.0 | 2.5 |
| 8 | 5 + 6 | 13 | 4.0 | 2.5 |
| 9 | 5 + 6 | 12 | 4.0 | 2.5 |
| 10 | 6 + 7 | 12 | 4.0 | 2.5 |
| 11 | 6 + 7 | 11 | 4.0 | 2.5 |
| 12 | 7 + 8 | 11 | 4.0 | 2.5 |
| 13 | 8 | 10 + 11 | 3.0 | 3.0 |
| 14 | 8 + 9 | 9 + 10 | 3.5 | 3.0 |

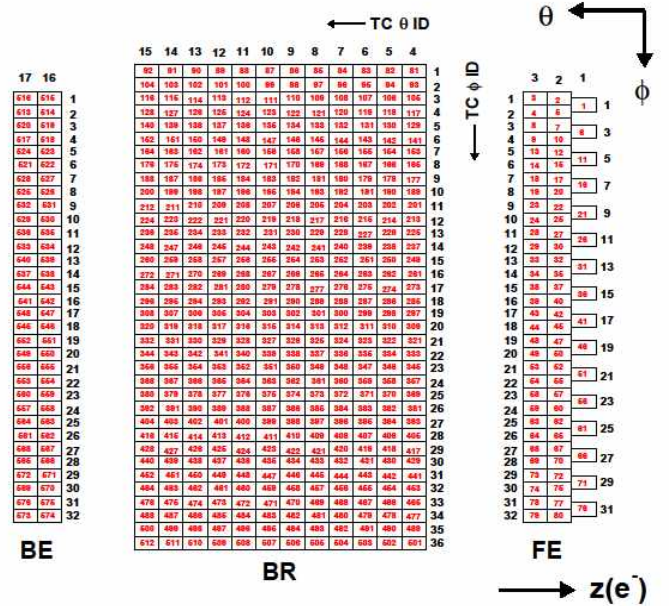

Fig. 3. The mapping layout from 576 Trigger Cells.

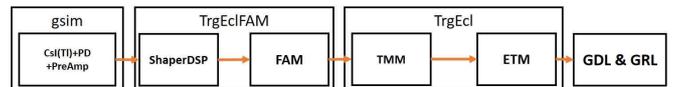

Fig. 4 TSim-ecl c++ object structure

The event timing is decided from the timing of the maximum energy deposited TC in an event because the timing resolution depends on pulse height. In the simulation study, the event timing resolution turned out to be 4 ns in one sigma level.

### IV. ECL TRIGGER SIMULATION

The performance of ECL trigger algorithms should be confirmed using ECL Trigger Simulation (TSim-ecl) package that is a C++ based program implemented in the Belle II Geant4-based analysis framework called Basf2. In order to develop and test appropriate trigger algorithms for high trigger efficiencies to diverse physics processes, TSim-ecl MC study was performed with same structure and function as actual ECL trigger system. Fig. 4 shows TSim-ecl object structure. Gsim is the simulation of the detector response with the GEANT4 package [4]. TrgEclFAM object processes ShaperDSP and FAM function. TrgEcl object processes the role of TMM and ETM.

TABLE II
PHYSICS TRIGGER EFFICIENCIES UPON 2-D AND 3-D BHABHA VETO LOGIC

| Sample | 2-D Logic (%) | 3-D Logic (%) |
|---|---|---|
| $e^+e^- \to \Upsilon(4S) \to B\bar{B}$ | 0.00 ± 0.00 | 0.00 ± 0.00 |
| Bhabha ($\theta_{bb} \geq 17°$) | 91.8 ± 0.27 | 91.8 ± 0.27 |
| ISR($e^+e^- \to \mu^+\mu^-$) | 0.20 ± 0.05 | 0.00 ± 0.00 |
| ISR($e^+e^- \to \pi^+\pi^-$) | 0.40 ± 0.06 | 0.10 ± 0.03 |
| $\tau \to$ generic decay | 0.50 ± 0.07 | 0.00 ± 0.00 |
| $\tau \to \mu\gamma$ | 2.00 ± 0.14 | 0.10 ± 0.03 |
| $\tau \to e\gamma$ | 3.70 ± 0.18 | 0.40 ± 0.06 |

In the SuperKEKB collider, Bhabha event process is a main physics background having not only the highest event rate but also similar event topology to low multiplicity physics process such as tau and initial state radiation (ISR) events. The 2-D logic has a weak point that does sometimes misidentify a low multiplicity process as a Bhabha event. In order to avoid such a case, we use the 3-D ($r - \theta - \phi$) back-to-back topology with tighter Bhabha veto trigger scheme than previous Belle. For realistic back-to-back event topology identification, we perform identifying isolated clusters in an event. Then we record the clusters upon back-to-back topology using theta and phi direction. After selecting two back-to-back clusters, we apply energy cut not only to the energy sum of two clusters (> 4 GeV) but also to the energy of each cluster (> 1 GeV).

Table II shows trigger efficiencies of various physics processes after Bhabha trigger veto using 2-D and 3-D Bhabha tagging logics. As we can see, the 3-D gives better performance while Bhabha tagging efficiencies are same in both 2-D and 3-D logic. In addition, we found that 3-D veto logic is better than 2-D logic to trigger tau lepton decays that are one of Belle II physics golden modes to test the Standard Model.

In order to test FPGA firmware logic implemented in ETM board, we studied the tendency of Bhabha trigger decision by comparing Bhabha MC sample to beam data taken from run #120 in Exp 3 luminosity run. Table III shows the ratio of number of identified events in each Bhabha trigger type to total number of events. It turns out that they are consistent in the beginning stage of data quality check and we confirm that current ETM Bhabha tagging firmware does work correct.

## V. CONCLUSION

From April 2018 the Belle II experiment using the SuperKEKB collider has started luminosity run to take beam data in order to search for the New Physics phenomena. The ECL trigger system has been well implemented with robust and flexible trigger scheme by using FADC and FPGA firmware combined architecture. The TSim-ecl simulation software has been developed in order to test and debug current ECL trigger algorithms implemented in ETM firmware. From TSim-ecl simulation study we conclude that physics trigger efficiencies on various physics processes by ECL trigger tagging are high enough. In addition 3-D Bhabha trigger logic does provide better selection of low multiplicity events that are crucial to Belle II physics potential for probing the New Physics. By using Tsim-ecl framework, we confirm that physics and Bhabha trigger logics are well implemented in ECL trigger system under Phase II operation.

TABLE III
THE RATIO OF BHABHA COMBINATIONS

| Bhabha trigger type | MC | Data (run #120) |
|---|---|---|
| 1 | 0.67 ± 0.00 | 0.62 ± 0.02 |
| 2 | 0.02 ± 0.00 | 0.03 ± 0.01 |
| 3 | 0.05 ± 0.00 | 0.08 ± 0.01 |
| 4 | 0.01 ± 0.00 | 0.02 ± 0.01 |
| 5 | 0.05 ± 0.00 | 0.08 ± 0.01 |
| 6 | 0.04 ± 0.00 | 0.07 ± 0.01 |
| 7 | 0.03 ± 0.00 | 0.05 ± 0.01 |
| 8 | 0.03 ± 0.00 | 0.06 ± 0.01 |
| 9 | 0.02 ± 0.00 | 0.03 ± 0.01 |
| 10 | 0.02 ± 0.00 | 0.04 ± 0.01 |
| 11 | 0.01 ± 0.00 | 0.02 ± 0.01 |
| 12 | 0.02 ± 0.00 | 0.02 ± 0.01 |
| 13 | 0.02 ± 0.00 | 0.03 ± 0.01 |
| 14 | 0.02 ± 0.00 | 0.01 ± 0.01 |